\documentclass[10pt,twocolumn,twoside]{IEEEtran}
\usepackage{algorithm}
\usepackage{algorithmic}
\usepackage{amsmath}
\usepackage{amssymb}
\usepackage{amsthm}
\usepackage{bbm}
\usepackage[final]{graphicx}
\usepackage{upgreek}

\DeclareMathOperator*{\argmin}{arg\,min}

\newtheorem{definitionCounter}{Definition}

\newtheorem{reproducing kernel}[propositionCounter]{Proposition}
\newtheorem{bandpass reconstruction}[theoremCounter]{Theorem}
\newtheorem{definiteness}[definitionCounter]{Definition}
\newtheorem{extended Bochner}[theoremCounter]{Theorem}

\title{Direct Reconstruction of Saturated Samples in Band-Limited OFDM Signals}

\author{Kyong Hwan Jin, Gain Kim, {\textit{Student Member, IEEE}}, Yusuf Leblebici, {\textit{Fellow, IEEE}}, Jong Chul Ye, {\textit{Senior Member, IEEE}}, Michael Unser, {\textit{Fellow, IEEE}}\thanks{Corresponding author: Michael Unser, {\'{E}}cole polytechnique f{\'{e}}d{\'{e}}rale de Lausanne, Biomedical Imaging Group, Station 17, CH--1015 Lausanne VD, Switzerland, E-mail: michael.unser@epfl.ch, URL: http://bigwww.epfl.ch/}\thanks{This work was supported in part by the European Research Council under Grant 692726 (H2020-ERC Project GlobalBioIm).}\thanks{K. H. Jin and M. Unser are with the Biomedical Imaging Group, {\'{E}}cole polytechnique f{\'{e}}d{\'{e}}rale de Lausanne, Switzerland (e-mail: kyonghwan.jin@gmail.com; michael.unser@epfl.ch).}
\thanks{G. Kim was with Microelectronic Systems Laboratory, {\'{E}}cole polytechnique f{\'{e}}d{\'{e}}rale de Lausanne, Lausanne, Switzerland, and is now with KAIST, Daejeon, Republic of Korea.(email: gikim@kaist.ac.kr).}
\thanks{Y. Leblebici is with Microelectronic Systems Laboratory, {\'{E}}cole polytechnique f{\'{e}}d{\'{e}}rale de Lausanne, Switzerland (email: yusuf.leblebici@epfl.ch).}\thanks{J. C. Ye is with Dept. of Bio and Brain Engineering and Dept. of Mathematical Sciences, KAIST, Daejeon, Republic of Korea. (e-mail: jong.ye@kaist.ac.kr).}}

\begin{document}
\maketitle\thispagestyle{empty}

%%%%%%%%%%%%%%%%%%%%%%%%%%%%%%%%%%%%%%%%%%%%%%%%%%%%%%%%%%%%%%%%%%%%%%%%%%%%%%%%
\begin{abstract}
Given a set of samples, a few of them being possibly saturated, we propose an efficient algorithm in order to cancel saturation while reconstructing band-limited signals. Our method satisfies a minimum-loss constraint and relies on sinc-related bases. It involves matrix inversion and is a direct, non-iterative approach. It consists of two main steps: (i) regression, to estimate the expansion coefficients of the signal model; (ii) interpolation, to restore an estimated value for those samples that are saturated. Because the proposed method is free from tuning parameters, it is hardware-friendly and we expect that it will be particularly useful in the context of orthogonal frequency-division multiplexing. There, the high peak-to-average power ratio of the transmitted signal results in a challenging decoding stage in the presence of saturation, which causes significant decoding errors due to the nonlinearity of amplifiers and receivers, ultimately resulting in band distortion and information loss. Our experiments on realistic simulations confirm that our proposed reconstruction of the saturated samples can significantly reduce transmission errors in modern high-throughput digital-communication receivers.
\end{abstract}

%\keywords{Clipping, orthogonal frequency-division multiplexing (OFDM), peak-to-average power ratio (PAPR).}
\normalfont

%%%%%%%%%%%%%%%%%%%%%%%%%%%%%%%%%%%%%%%%%%%%%%%%%%%%%%%%%%%%%%%%%%%%%%%%%%%%%%%%
\section{Introduction}\label{sec: Introduction}

%Innovative solutions are needed to quench the exponential growth of the data traffic related to mobile devices, combined with the increasing demand for large-scale data processing in cloud systems and the expansion of the communication bandwidth required for processor-to-processor and processor-to-memory connections, all resulting in ever-higher wireless communication data rates under a finite-energy budget. Fortunately, orthogonal frequency-division multiplexing (OFDM) offers high spectral efficiency and robustness to multi-path distortion~\cite{bingham1990multicarrier}; it has been used for the IEEE 802.11a,g,n,ac standards of wireless local area network and for the 4G-LTE standards.

Innovative solutions such as orthogonal frequency-division multiplexing (OFDM) are needed to quench the exponential growth of the data traffic related to mobile devices~\cite{bingham1990multicarrier}. One major drawback of OFDM, however, is its high peak-to-average power ratio (PAPR) \cite{jiang2008overview}. As OFDM subdivides the frequency bands into several subcarriers, the amplitude of the transmitted signals has a large dynamic range. This results in band distortions when the input exceeds the linear region of the power amplifier. While the band-distortion problem can sometimes be resolved by highly linear power amplifiers, their deployment typically reduces the power efficiency of the hardware and increases costs.

PAPR-reduction techniques have been proposed such as clipping~\cite{o1995envelope,ochiai2000performance}, phase optimization~\cite{tarokh2000computation}, tone reservation/injection~\cite{yoo2006novel}, partial transmission sequence~\cite{muller1997ofdm}, and selective mapping~\cite{bauml1996reducing}. The last two methods reduce the PAPR at the transmitter side while distortions are compensated by transmitting additional bits through a side channel.

An alternative PAPR-reduction approach that makes the economy of a side channel is to detect the occurrence of saturated samples and then substitute them with estimates based on prior assumptions. For instance, Marvasti {\textit{et al.}}~\cite{saeedi2002clipping,Marvasti2001,Marvasti1991} make the assumption that the underlying OFDM signal is band-limited and then devise an iterative scheme to reconstruct the saturated values\footnote{The reconstruction of band-limited signals is a classical problem~\cite{jerri1977shannon,Feichtinger1995,Aldroubi2001b}; it can be traced back to the Papoulis-Gerchberg algorithm~\cite{Gerchberg1974,Papoulis1975}}. More recently, a Bayesian reconstruction that takes advantage of the sparse nature of the underlying signals and combines it with compressed sensing was proposed in~\cite{al2012peak}. Extending a single OFDM reconstruction to the multiuser OFDM reconstructions that take place in modern single-input multiple-output systems was further proposed in~\cite{ali2014receiver}. However, the aforementioned iterative solutions~\cite{saeedi2002clipping,al2012peak,ali2014receiver} are not hardware-friendly and lack in robustness, being highly sensitive to the tuning of their parameters.

In~\cite{bibi2016equation}, the pseudo-inversion of a partial discrete-Fourier-transform matrix constructed with the locations of saturation (time) and so-called ``reliable components'' (frequency) is used to estimate the values of the true signal. This requires one to assume that the reliable components are known to every underlying OFDM receiver. In practice, however, there is no such information at the receiver and the reliable components have to be guessed.

In this paper, we propose a direct, non-iterative algorithm that compensates for the saturation of band-limited OFDM signals, thus attenuating band distortions and resulting in an improved bit-error-ratio of the transmitted bit streams. We derive the theory of minimum-norm reconstructions with respect to sampling consistency and reproducing kernel in the space of band-limited functions. The theory reveals that our approach gives a solution that is unique, freed from the uncertainty that was tied to the reliable-component method of~\cite{bibi2016equation}. Our algorithm consists of two steps. The first step (regression) amounts to a matrix inversion and the second step (interpolation) to a simple matrix-vector multiplication. The fact that our algorithm is direct makes it very affordable in practice since this leads to a saving of critical resources such as computing cycles. Moreover, it turns out that our algorithm is able to obtain a coefficient estimate from fewer samples than is required by ideal sampling. This property leads to a further reduction of the computational effort in case of long signals. Finally, our algorithm shows stable performance with respect to its tuning parameters in a variety of experimental scenarios.

This paper is organized as follows: In Section~\ref{sec: Theory}, we provide a fundamental theorem for nonuniform interpolation under the assumption of band-limitedness. In Section~\ref{sec: Proposed Method}, we describe our algorithm and its regression and interpolation steps. In Section~\ref{sec: Results}, we demonstrate the capability of our algorithm to recover the saturated samples of band-limited signals in numerical and realistic OFDM simulations.

%%%%%%%%%%%%%%%%%%%%%%%%%%%%%%%%%%%%%%%%%%%%%%%%%%%%%%%%%%%%%%%%%%%%%%%%%%%%%%%%
\section{Theory}\label{sec: Theory}

%%%%%%%%%%%%%%%%%%%%%%%%%%%%%%%%%%%%%%%%%%%%%%%%%%%%%%%%%%%%%%%%%%%%%%%%%%%%%%%%
\subsection{Terminology}\label{sec: Terminology}
%%------------------------
\begin{figure}
\begin{center}
\includegraphics[trim = 65mm 85mm 60mm 60mm,clip=true,width=3.7in ]{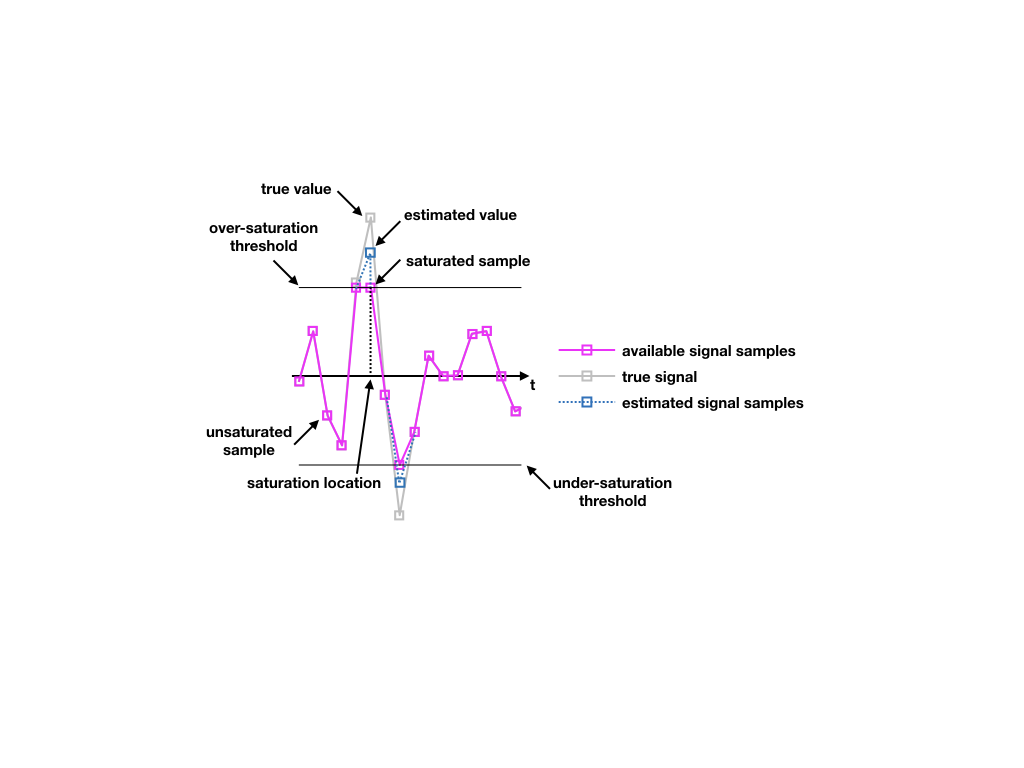}
\caption{\label{fig: terminology}Terminology.}
\end{center}
\end{figure}
%%------------------------
We illustrate in Figure~\ref{fig: terminology} the terminology that we shall use throughout this paper.
\begin{itemize}
\item True signal: analog, unknown signal.
\item Sample: a (location, value) pair. We assume that all locations are quantized on a regular grid.
\item True value: value of a sample of the true signal.
\item Unsaturated sample: sample of the true signal, taken within the linear range of the amplifier. The values of unsaturated samples coincide with true values.
%\item Under-saturation threshold: lower bound of the linear domain of the amplifier.
%\item Over-saturation threshold: upper bound of the linear domain of the amplifier.
\item Saturated sample: the (location, under-saturation threshold) or (location, over-saturation threshold) pair that is returned by the hardware whenever the signal exceeds the linear domain of the amplifier. The detection of saturated samples is achieved by monitoring the sample values; no side channel is required.
\item Estimated value: outcome of our approach.
\end{itemize}

%%%%%%%%%%%%%%%%%%%%%%%%%%%%%%%%%%%%%%%%%%%%%%%%%%%%%%%%%%%%%%%%%%%%%%%%%%%%%%%%
\subsection{Assumptions}\label{sec: Assumptions}
The strategy for this work is to fit unsaturated samples with a band-limited curve and then determine (reconstruct) estimated values by resampling the fit at the locations where saturation did occur. When stated in full generality, this is an ill-posed problem---think of the case when there are but saturated samples. To make the solution tractable, we need additional hypotheses. We consider here the conjunction of three assumptions that are realistic in the OFDM context: (i) the saturated samples are few and their spacing lacks regularity; (ii) the unsaturated samples have low noise; (iii) the true signal is band-limited.

%%%%%%%%%%%%%%%%%%%%%%%%%%%%%%%%%%%%%%%%%%%%%%%%%%%%%%%%%%%%%%%%%%%%%%%%%%%%%%%%
\subsection{Development}\label{sec: Development}
We are going to show that our reconstruction problem admits a closed-form solution that is related to the kernel methods used in machine learning~\cite{smale2004, Smale2005}. The enabling property is that the Shannon space of band-limited functions can be endowed with the structure of a reproducing-kernel Hilbert space~\cite{Yao1967,Nashed1991}. Within this setting, we shall derive an extension of the traditional sampling theory to recover bandpass functions from nonuniform samples. The nominal band of the signal is a (not necessarily connected) set $B_{0}\subset{\mathbb{R}}$ that is symmetric around the origin and whose measure (or bandwidth) is finite. We define the corresponding space of bandpass functions as
\begin{equation}
\label{eq: HB0}{\mathcal{H}}_{B_{0}}=\{f\in L_{2}({\mathbb{R}}):\hat{f}(\omega)=0,\forall\omega\not\in B_{0}\},
\end{equation}
where $\hat{f}(\omega):={\mathcal{F}}\{f\}(\omega)=\int_{{\mathbb{R}}}\,f(t)\,{\mathrm{e}}^{-{\mathrm{j}}\,\omega\,t}\,{\mathrm{d}}t$ is the Fourier transform of the continuous-domain signal $f$ and $L_{2}({\mathbb{R}})$ is Lebesgue's space of finite-energy functions. With this notation, ${\mathcal{H}}_{(-\uppi,\uppi)}$ coincides with Shannon's space of band-limited functions~\cite{Unser2000}, which is spanned\footnote{The dot notation ``$\cdot$'' is a placeholder for mute variables, oftentimes temporal ones.} by the orthonormal basis $\{{\mathrm{sinc}}(\cdot-n)\}_{n\in{\mathbb{Z}}}$. The main extension here is that the spectral band $B_{0}$ can be arbitrary and that we do not require the existence of a basis for ${\mathcal{H}}_{B_{0}}$, but rather a frame~\cite{Christensen2003}. More precisely, letting $S$ be an adequate sampling set, this frame is an over-complete family $\{\phi(\cdot-\tau)\}_{\tau\in S}$ of functions that span ${\mathcal{H}}_{B_{0}}$.

%%%%%%%%%%%%%%%%%%%%%%%%%%%%%%%%%%%%%%%%%%%%%%%%%%%%%%%%%%%%%%%%%%%%%%%%%%%%%%%%
\subsection{Underlying Hilbert-Space Structure}\label{sec: Underlying Hilbert-Space Structure}
%%------------------------
\begin{figure}
\begin{center}
\includegraphics[width=8cm]{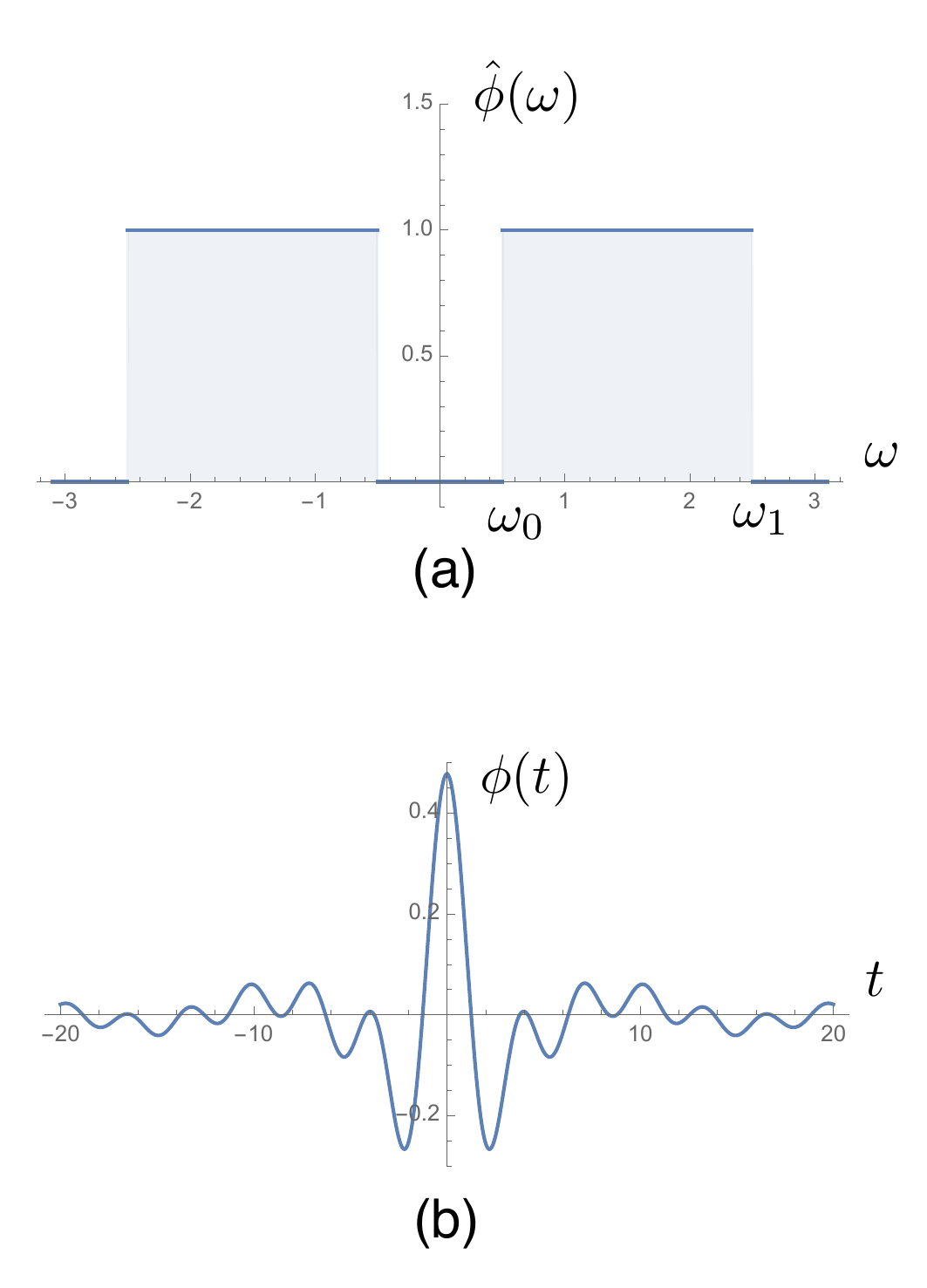}
\caption{\label{fig: bandpass}Bandpass scenario. Top: frequency response of the ideal bandpass filter. Bottom: bandpass reproducing kernel in the temporal domain.}
\end{center}
\end{figure}
%%------------------------
We now prove that ${\mathcal{H}}_{B_{0}}$ is a Hilbert space by identifying its reproducing kernel. To that end, we consider the indicator function
\begin{equation}
\label{eq: Fourier phi}\hat{\phi}(\omega)={\mathbbm{1}}_{B_{0}}(\omega)=\left\{\begin{array}{ll}1,&\omega\in B_{0}\\0,&\omega\not\in B_{0}\end{array}\right.
\end{equation}
of the set $B_{0}$ in the frequency-domain, along with its inverse Fourier transform
\begin{equation}
\label{eq: phi}\phi(t)={\mathcal{F}}^{-1}\{\hat{\phi}\}(t)=\frac{1}{2\,\uppi}\,\int_{{\mathbb{R}}}\,\hat{\phi}(\omega)\,{\mathrm{e}}^{{\mathrm{j}}\,\omega\,t}\,{\mathrm{d}}\omega.
\end{equation}
Both are illustrated in Figure~\ref{fig: bandpass}.

Given the kernel $\phi$, we then define the convolution operator ${\mathcal{R}}:L_{2}\rightarrow{\mathcal{H}}_{B_{0}}$ as
\begin{equation}
{\mathcal{R}}\{f\}(t)=\left(\phi*f\right)(t)=\int_{{\mathbb{R}}}\,\phi(t-\tau)\,f(\tau)\,{\mathrm{d}}\tau,
\end{equation}
which acts as an ideal bandpass filter. By invoking properties of the ideal filter, we then show idempotence
\begin{equation}
\label{eq: idempotence}\forall g\in{\mathcal{H}}_{B_{0}}:{\mathcal{R}}\{g\}=g
\end{equation}
and orthogonality of the error
\begin{equation}
\label{eq: error orthogonality}\left\langle f-{\mathcal{R}}\{f\},g\right\rangle=\frac{1}{2\,\uppi}\left\langle\left(1-\hat{\phi}\right)\,\hat{f},\hat{g}\right\rangle=0
\end{equation}
for any $f\in L_{2}({\mathbb{R}})$, which proves that ${\mathcal{R}}$ is the orthogonal projector of $L_{2}$ onto ${\mathcal{H}}_{B_{0}}$. Likewise, we readily verify that ${\mathcal{R}}$ is a unitary ${\mathcal{H}}_{B_{0}}\rightarrow{\mathcal{H}}_{B_{0}}$ operator when ${\mathcal{H}}_{B_{0}}$ is equipped with the usual $L_{2}$ inner product
\begin{equation}
\label{eq: H-inner product}\left\langle f,g\right\rangle_{{\mathcal{H}}_{B_{0}}}=\left\langle f,g\right\rangle:=\int_{{\mathbb{R}}}\,f(t)\,g(t)\,{\mathrm{d}}t.
\end{equation}
This suggests that the space ${\mathcal{H}}_{B_{0}}$ is its own topological dual and that ${\mathcal{R}}$ is the corresponding Riesz map whose generalized impulse response is precisely the sought-after reproducing kernel.

\begin{reproducing kernel}[Identification of the reproducing kernel]\label{prop: reproducing kernel}Let $\phi(t)=\frac{1}{2\,\uppi}\,\int_{B_{0}}\,{\mathrm{e}}^{{\mathrm{j}}\,\omega\,t}\,{\mathrm{d}}\omega$. Then, the bivariate function $h(t,\tau)=\phi(t-\tau)$ is such that, for any fixed $\tau\in{\mathbb{R}}$ and $f\in{\mathcal{H}}_{B_{0}}$,
\begin{equation}
\label{eq: h=phi}t\mapsto h(t,\tau)=\phi(t-\tau)\in{\mathcal{H}}_{B_{0}}
\end{equation}
\begin{equation}
\label{eq: <f,h>=f}\left\langle f,h(\cdot,\tau)\right\rangle=\left\langle f,\phi(\cdot-\tau)\right\rangle=f(\tau).
\end{equation}
Hence, $h:{\mathbb{R}}\times{\mathbb{R}}\rightarrow{\mathbb{R}}$ is the reproducing kernel of the Hilbert space ${\mathcal{H}}_{B_{0}}$ equipped with the $L_{2}$ inner product.
\begin{proof}
By definition, the reproducing kernel of a Hilbert space $({\mathcal{H}},\left\langle\cdot,\cdot\right\rangle_{{\mathcal{H}}})$ is the unique bivariate function $h(t,\tau)$ such that $h(\cdot,\tau_{0})\in{\mathcal{H}}$ and such that $f(\tau_{0})=\left\langle f,h(\cdot,\tau_{0})\right\rangle_{{\mathcal{H}}}$ for all $f\in{\mathcal{H}}$ and any fixed $\tau_{0}$~\cite{Aronszajn1950}. In the present scenario, the kernel is shift-invariant and the two conditions are established by considering inclusion and reproduction.
\begin{itemize}
\item Inclusion property: The Fourier transform of $\phi(\cdot-\tau_{0})$ is ${\mathrm{e}}^{-{\mathrm{j}}\,\omega\,\tau_{0}}\,\hat{\phi}(\omega)$, $\omega\in{\mathbb{R}}$. This is a finite-energy function supported in $B_{0}$, which proves that $\phi(\cdot-\tau_{0})\in{\mathcal{H}}_{B_{0}}$.
\item Reproduction property: The function $\hat{\phi}={\mathbbm{1}}_{B_{0}}$ is symmetric and real-valued, so that the same holds true for its inverse Fourier transform $\phi$. Consequently, the central part of~(\ref{eq: <f,h>=f}) is equivalent to the convolution product $\left(f*h\right)(\tau)=\left\langle f,\phi(\tau-\cdot)\right\rangle$. Its Fourier-domain counterpart is $\hat{\phi}\,\hat{f}=\hat{f}$ since $\hat{\phi}$ is the indicator function of $B_{0}$ and $\hat{f}(\omega)=0$ for all $\omega\not\in B_{0}$. The conclusion ensues.
\end{itemize}
\end{proof}
\end{reproducing kernel}

%%%%%%%%%%%%%%%%%%%%%%%%%%%%%%%%%%%%%%%%%%%%%%%%%%%%%%%%%%%%%%%%%%%%%%%%%%%%%%%%
\subsection{Minimum-Norm Reconstruction}\label{sec: Minimum-Norm Reconstruction}
%%------------------------
\begin{figure*}
\begin{center}
% \hline
% \includegraphics[width=6.5in]{fig03}
\includegraphics[trim = 35mm 75mm 35mm 75mm,clip=true,width=6.25in]{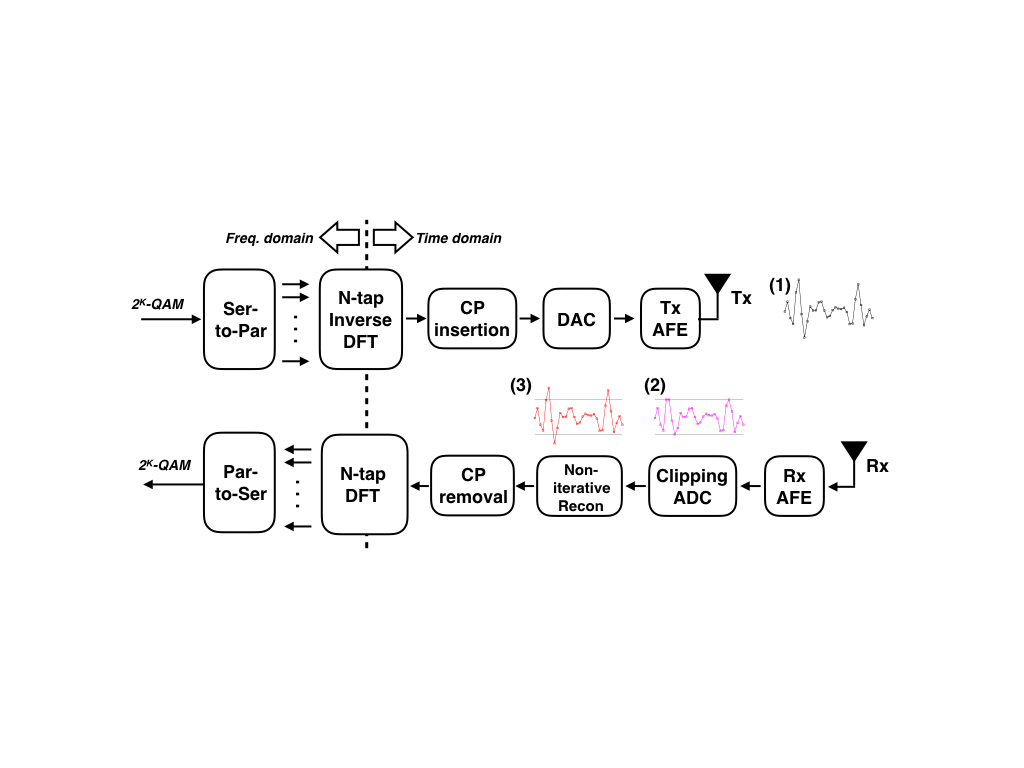}
% \hline
\caption{\label{fig: flowchart}Flowchart in the temporal domain. Tx: transmitter. Rx: receiver. ADC: analog-to-digital converter. Non-iterative Recon: our interpolation method.}
\end{center}
\end{figure*}
%%------------------------
Since the function $f$ is an infinite-dimensional entity, its reconstruction from a finite number $N$ of (possibly nonuniform) samples $f(t_{1} ),\ldots,f(t_{N})$ is ill-posed in general. We regularize the problem by searching for the minimum-norm solution within the space ${\mathcal{H}}_{B_{0}}$ of bandpass functions. The remarkable outcome of our formulation is that this solution is always well defined, irrespectively of the sampling locations $t_{n}$. Furthermore, it can be found efficiently by solving a linear system of equations.
\begin{bandpass reconstruction}[Bandpass signal reconstruction]\label{thm: bandpass reconstruction}Let $S=\{(t_{n},f_{n})\}_{n=1}^{N}$ be a list of real-valued samples. The minimum-norm interpolant of $S$ in ${\mathcal{H}}_{B_{0}}$, in the sense that
\begin{equation}
\label{eq: fint}f_{{\mathrm{int}}}=\argmin_{\stackrel{f\in{\mathcal{H}}_{B_{0}}}{\forall n\in[1\ldots N]:f(t_{n})=f_{n}}}\left\|f\right\|_{L_{2}({\mathbb{R}})}^{2},
\end{equation}
is given by
\begin{equation}
\label{eq: fint(.)}t\mapsto f_{{\mathrm{int}}}(t)=\sum_{n=1}^{N}\,\alpha_{n}\,\phi(t-t_{n}),
\end{equation}
where $\phi(t)=\frac{1}{2\,\uppi}\,\int_{B_{0}}\,{\mathrm{e}}^{{\mathrm{j}}\,\omega\,t}\,{\mathrm{d}}\omega$ and $\mbox{{\boldmath$\upalpha$}}=(\alpha_{1},\ldots,\alpha_{N})={\mathbf{R}}^{-1}\,{\mathbf{f}}$, with ${\mathbf{f}}=(f_{1},\ldots,f_{N})$ and $\left[{\mathbf{R}}\right]_{m,n}=r_{m,n}=\phi(t_{m}-t_{n})$. Moreover, ${\mathbf{R}}$ is guaranteed to be invertible whenever the sample locations $t_{n}$ are distinct, which ensures that the solution is unique and well-defined.
\begin{proof}
First, we define the basis fonctions $\phi_{n}=\phi(\cdot-t_{n})\in{\mathcal{H}}_{B_{0}}$ for $n\in[1\ldots N]$. They are such that
\begin{equation}
\label{eq: f(tm)}f(t_{n})=\left\langle\phi_{n},f\right\rangle
\end{equation}
\begin{equation}
\label{eq: <phim,phin>}\left\langle\phi_{m},\phi_{n}\right\rangle=\phi(t_{m}-t_{n})=r_{m,n},
\end{equation}
as a direct consequence of~(\ref{eq: <f,h>=f}). We then decompose the space ${\mathcal{H}}_{B_{0}}={\mathcal{V}}\oplus{\mathcal{V}}^{\perp}$ as the direct sum of ${\mathcal{V}}={\mathrm{span}}\{\phi_{n}\}_{n=1}^{N}$ and its orthogonal complement ${\mathcal{V}}^{\perp}$ in in ${\mathcal{H}}_{B_{0}}$, with
\begin{equation}
\label{eq: Vperp}{\mathcal{V}}^{\perp}=\{f\in{\mathcal{H}}_{B_{0}}:\left\langle\phi_{n},f\right\rangle=0,\forall n\in[1\ldots N]\}.
\end{equation}
Consequently, any $f\in{\mathcal{H}}_{B_{0}}$ has the unique decomposition $f=u+v$ with $u\in{\mathcal{V}}^{\perp}$ and $v\in{\mathcal{V}}$. By construction, we also have the Pythagorean relation $\left\|f\right\|_{L_{2}}^{2}=\left\|u\right\|_{L_{2}}^{2}+\left\|v\right\|_{L_{2}}^{2}$, due to the equivalence of the ${\mathcal{H}}_{B_{0}}$ norm and the $L_{2}$ norm. The condition $u\in{\mathcal{V}}^{\perp}$ is equivalent to $u(t_{n})=\left\langle\phi_{n},u\right\rangle=0$ for $n\in[1\ldots N]$ which, in reason of the minimum-norm requirement, leads to $f_{{\mathrm{int}}}=u_{{\mathrm{int}}}\in{\mathcal{V}}$.

Finally, we apply the interpolation constraint to the parametric representation of $f_{{\mathrm{int}}}$ which, owing to~(\ref{eq: f(tm)}) and~(\ref{eq: <phim,phin>}), results in the set of linear equations given by
\begin{eqnarray}
\nonumber f_{{\mathrm{int}}}(t_{m})&=&\left\langle\phi_{m},f_{{\mathrm{int}}}\right\rangle=\sum_{n=1}^{N}\,\alpha_{n}\,\left\langle\phi_{m},\phi_{n}\right\rangle\\
\label{eq: fint(tm)}&=&\sum_{n=1}^{N}\,\alpha_{n}\,\phi(t_{m}-t_{n})=\sum_{n=1}^{N}\,r_{m,n}\,\alpha_{n}
\end{eqnarray}
in terms of $\alpha_{n}$, with $n\in[1\ldots N]$. The important additional ingredient is the positive definiteness of $\phi$, which follows from the extension of Bochner's theorem given in Theorem~\ref{thm: extended Bochner}.
\end{proof}
\end{bandpass reconstruction}

\begin{definiteness}[Definiteness]\label{def: definiteness}A function $\phi:{\mathbb{R}}\rightarrow{\mathbb{R}}$ is said to be positive semidefinite if
\begin{equation}
\label{eq: +semidefinite}\sum_{n=1}^{N}\,\sum_{m=1}^{N}\,z_{m}\,\phi(t_{m}-t_{n})\,z_{n}\geq0
\end{equation}
for every $t_{1},\ldots,t_{N}\in{\mathbb{R}}$, $z_{1},\ldots,z_{N}\in{\mathbb{R}}$, and any positive integer $N$. Likewise, $\phi$ is said to be positive definite if the nonnegative relation in~(\ref{eq: +semidefinite}) can be replaced by a positive relation for any non-vanishing ${\mathbf{z}}=(z_{1},\ldots,z_{N})\in{\mathbb{R}}^{N}\setminus\{{\mathbf{0}}\}$ and distinct $t_{n}\in{\mathbb{R}}$.
\end{definiteness}

\begin{extended Bochner}[Extended Bochner's theorem~\cite{Wendland2005}]\label{thm: extended Bochner}A continuous function $\phi:{\mathbb{R}}\rightarrow{\mathbb{R}}$ is positive semidefinite if and only if it is the Fourier transform of a positive Borel measure, so that $\phi(t)=\frac{1}{2\,\uppi}\,\int_{{\mathbb{R}}}\,\hat{\phi}(\omega)\,{\mathrm{e}}^{{\mathrm{j}}\,\omega\,t}\,{\mathrm{d}}\omega$, with $\hat{\phi}(\omega)\geq0$ for all $\omega\in{\mathbb{R}}$. Moreover, it is positive definite if and only if the carrier of $\hat{\phi}$ is nonzero; in other words, if and only if there is a measurable set $E$ such that $\hat{\phi}(\omega)\ne0$ for all $\omega\in E$.
\end{extended Bochner}

In our case, the last condition is met trivially since the carrier of $\hat{\phi}$ is precisely $E=B_{0}$. Finally, the positive definiteness of $\phi$ implies the invertibility of the symmetric matrix ${\mathbf{R}}$ for any given choice of distinct sampling locations $t_{1},\ldots,t_{N}$.

%%%%%%%%%%%%%%%%%%%%%%%%%%%%%%%%%%%%%%%%%%%%%%%%%%%%%%%%%%%%%%%%%%%%%%%%%%%%%%%%
\subsection{Particular Cases}\label{sec: Particular Cases}
For the Nyquist band, when $B_{0}=(-\uppi,\uppi)$, we have that $\phi$ is the {\textit{sinus cardinalis}} $\phi(t)={\mathrm{sinc}}\,(t)=\frac{\sin(\uppi\,t)}{\uppi\,t}$ for $t\in{\mathbb{R}}$. This yields the solution $f_{{\mathrm{int}}}(t)=\sum_{n=1}^{N}\,\alpha_{n}\,{\mathrm{sinc}}(t-t_{n})$, which is reminiscent of Shannon's sampling expansion. The caveat, of course, is that the optimal expansion coefficients $\alpha_{n}$ are usually distinct from $f_{n}=f(t_{n})$, unless the sampling locations $t_{n}$ build a regular grid.

In the bandpass scenario, we have that
\begin{equation}
\label{eq: B0}B_{0}=(-\omega_{1},-\omega_{0}]\cup[\omega_{0},\omega_{1})
\end{equation}
with $0\leq\omega_{0}<\omega_{1}<\infty$. This corresponds to the reproducing kernel
\begin{equation}
\phi(t)=\frac{2\,\uppi}{\omega_{1}}\,{\mathrm{sinc}}\frac{\omega_{1}\,t}{2\,\uppi}-\frac{2\,\uppi}{\omega_{0}}\,{\mathrm{sinc}}\frac{\omega_{0}\,t}{2\,\uppi}.
\end{equation}
The relevant functions for this construction are illustrated in Figure~\ref{fig: bandpass}.

%%%%%%%%%%%%%%%%%%%%%%%%%%%%%%%%%%%%%%%%%%%%%%%%%%%%%%%%%%%%%%%%%%%%%%%%%%%%%%%%
\section{Proposed Method}\label{sec: Proposed Method}
In this section, we describe the method to obtain $\mbox{{\boldmath{$\upalpha$}}}$ in Theorem~\ref{thm: bandpass reconstruction} in the context of few saturated samples coming in an irregular sequence. The overall OFDM system is illustrated in Figure~\ref{fig: flowchart}. The input and output bit streams are encoded with $2^{K}$ quadrature amplitude modulation (QAM) and the encoded signal is described by its frequency components. The multi-carrier OFDM signal is generated by an inverse discrete Fourier transform (IDFT). The OFDM signal is transmitted in the time domain. The effect of saturation (clipping) is modeled by a sampling operation combined with an analog-to-digital converter (ADC). The under-saturation threshold $T_{0}$ and the over-saturation threshold $T_{1}$ depend on the specifications of the ADC; once the unknown signal $f$ is saturated and sampled, the accessible values are
\begin{equation}
f_{{\mathrm{s}}}(t_{n})=\left\{\begin{array}{ll}T_{0},&f(t_{n})\leq T_{0}\\f_{n},&T_{0}<f(t_{n})<T_{1}\\T_{1},&T_{1}\leq f(t_{n}).\end{array}\right.
\end{equation}

It is now crucial to realize that the theory developed in Section~\ref{sec: Theory} does not require any specific temporal ordering of the batch of $M=N+\left(M-N\right)$ samples we have at our disposal. We take advantage of this opportunity to re-index them and partition them in two groups. The first group is made of $N$ unsaturated samples indexed by $n\in[1\ldots N]$ such that $f_{{\mathrm{s}}}(t_{n})=f_{n}\not\in\{T_{0},T_{1}\}$. Meanwhile, the second group contains the remaining $\left(M-N\right)$ saturated samples indexed by $n\in[N+1\ldots M]$ such that $f_{{\mathrm{s}}}(t_{n})\in\{T_{0},T_{1}\}$. By notational convention, we collect the unsaturated OFDM signal values in the vector ${\mathbf{y}}=(f_{1},\ldots,f_{N})\in{\mathbb{R}}^{N}$.

%\fbox{\parbox{3in}{{\textbf{PT: I'm not sure this is important.}} Note that the basis function is circularly shifted around the boundaries because the OFDM signal is assumed to be acquired on a grid and finite in discrete domain.}}

%%%%%%%%%%%%%%%%%%%%%%%%%%%%%%%%%%%%%%%%%%%%%%%%%%%%%%%%%%%%%%%%%%%%%%%%%%%%%%%%
\subsection{Regression}\label{sec: Regression}
%%------------------------
\begin{figure*}
\begin{center}
\begin{equation}
\label{eq: regression}\underbrace{\left(\begin{array}{c}\alpha_{1}\\\vdots\\\alpha_{N}\end{array}\right)}_{\mbox{{\boldmath{$\upalpha$}}}}=\underbrace{\left(\begin{array}{cccc}\phi(t_{1}-t_{1})&\phi(t_{1}-t_{2})&\cdots&\phi(t_{1}-t_{N})\\\phi(t_{2}-t_{1})&\phi(t_{2}-t_{2})&\cdots&\phi(t_{2}-t_{N})\\\vdots&\vdots&\ddots&\vdots\\\phi(t_{N}-t_{1})&\phi(t_{N}-t_{2})&\cdots&\phi(t_{N}-t_{N})\end{array}\right)^{-1}}_{{\mathbf{R}}^{-1}}\,\underbrace{\left(\begin{array}{c}f_{1}\\\vdots\\f_{N}\end{array}\right)}_{{\mathbf{y}}}
\end{equation}
%\rule{\textwidth}{1pt}
\end{center}
\end{figure*}
%%------------------------
Applying~(\ref{eq: fint(tm)}) to the group of all $N$ unsaturated samples, we derive that $\mbox{{\boldmath{$\upalpha$}}}={\mathbf{R}}^{-1}\,{\mathbf{y}}$. This regression step is written explicitly in~(\ref{eq: regression}). Due to the symmetric structure of $B_{0}$ assumed in~(\ref{eq: B0}), the matrix to invert is symmetric, too, which facilitates computations. Bad conditioning can be further avoided by an $\varepsilon$-parametric regularization~\cite{kybic2002generalized,kybic2002generalized_2}, as in
\begin{equation}
\label{eq: regularization}\mbox{{\boldmath{$\upalpha$}}}=\left({\mathbf{R}}+\varepsilon\,{\mathbf{I}}\right)^{-1}\,{\mathbf{y}}.
\end{equation}

%%%%%%%%%%%%%%%%%%%%%%%%%%%%%%%%%%%%%%%%%%%%%%%%%%%%%%%%%%%%%%%%%%%%%%%%%%%%%%%%
\subsection{Interpolation}\label{sec: Interpolation}
%%------------------------
\begin{figure*}
\begin{center}
\begin{equation}
\label{eq: interpolation}\left(\begin{array}{c}f_{{\mathrm{int}}}(t_{N+1})\\\vdots\\f_{{\mathrm{int}}}(t_{M})\end{array}\right)=\underbrace{\left(\begin{array}{cccc}\phi(t_{N+1}-t_{1})&\phi(t_{N+1}-t_{2})&\cdots&\phi(t_{N+1}-t_{N})\\\phi(t_{N+2}-t_{1})&\phi(t_{N+2}-t_{2})&\cdots&\phi(t_{N+2}-t_{N})\\\vdots&\vdots&\ddots&\vdots\\\phi(t_{M}-t_{1})&\phi(t_{M}-t_{2})&\cdots&\phi(t_{M}-t_{N})\end{array}\right)}_{{\mathbf{E}}}\,\underbrace{\left(\begin{array}{c}\alpha_{1}\\\vdots\\\alpha_{N}\end{array}\right)}_{\mbox{{\boldmath{$\upalpha$}}}}
\end{equation}
\rule{\textwidth}{1pt}
\end{center}
\end{figure*}
%%------------------------
The joint application of~(\ref{eq: fint(.)}) to the group of all $\left(M-N\right)$ saturated samples allows us to determine their estimated values. This interpolation step is written explicitly in~(\ref{eq: interpolation}). Since it involves a matrix ${\mathbf{E}}\in{\mathbb{R}}^{\left(M-N\right)\times N}$, the total computational effort is ${\mathcal{O}}(N^{3}+M\,N-N^{2})={\mathcal{O}}(N^{3})$, in reason of our assumption that $N\gg\left(M-N\right)$.

%%%%%%%%%%%%%%%%%%%%%%%%%%%%%%%%%%%%%%%%%%%%%%%%%%%%%%%%%%%%%%%%%%%%%%%%%%%%%%%%
\subsection{Shortcomings}\label{sec: Shortcomings}
In the presence of regularization ({\textit{i.e.}}, $\varepsilon\neq0$), the equality $f_{{\mathrm{int}}}(t_{n})=f_{n}$ is not necessarily valid anymore if we use~(\ref{eq: regularization}) to determine $\mbox{{\boldmath{$\upalpha$}}}$ in~(\ref{eq: fint(.)}). Thus, the combined list of samples $\{(t_{n},f_{n})\}_{n=1}^{N}\cup\{(t_{n},f_{{\mathrm{int}}}(t_{n}))\}_{n=N+1}^{M}$ is not consistent with our band-pass model. Despite this discrepancy, and for sheer convenience, we shall disregard this state of affairs in the interest of simplicity and reconstruct only the saturated samples, leaving the unsaturated ones pristine.

Likewise, we do not attempt to confirm whether the estimated values do indeed exceed the saturation thresholds. While special configurations of saturated values exist where ringing can mislead our method to return estimated values within the interval $(T_{0},T_{1})$, these configurations never arise in practice. Solving for these rare cases would require linear-programming methods, which come at an unacceptable cost-benefit tradeoff.

%%%%%%%%%%%%%%%%%%%%%%%%%%%%%%%%%%%%%%%%%%%%%%%%%%%%%%%%%%%%%%%%%%%%%%%%%%%%%%%%
\subsection{Sliding Window}\label{sec: Sliding Window}
%%------------------------
\begin{algorithm}[t]
\begin{center}
\begin{algorithmic}[1]
\REQUIRE{$\{(t_{n},f_{{\mathrm{s}}}(t_{n}))\}_{n=1}^{M}$, $N$, $\varepsilon$}
\\\COMMENT{Find unsaturated samples, $K$ of them}
\STATE{$K=0$; $m=M$}
\FOR{$n=1$ \TO$M$}
\IF{$f_{{\mathrm{s}}}(t_{n})\in\{T_{0},T_{1}\}$}
\STATE{$(\tau_{m},\varphi_{m})=(t_{n},f_{{\mathrm{s}}}(t_{n}))$; $m\leftarrow m-1$}
\ELSE\STATE{$K\leftarrow K+1$; $(\tau_{K},\varphi_{K})=(t_{n},f_{{\mathrm{s}}}(t_{n}))$}
\ENDIF
\ENDFOR
\IF{$K<N$}\STATE{Return with failure (saturation too dense)}\ENDIF
\\\COMMENT{Estimate the $\left(M-K\right)$ saturated samples}
\FOR{$k=K+1$ \TO$M$}
%\FOR{$m=1$ \TO$K$}
\STATE{$\forall m\in[1\ldots K]:u_{m}=\left|\tau_{m}-\tau_{k}\right|$}
%\ENDFOR
\STATE{Find a permutation matrix ${\mathbf{P}}\in\{0,1\}^{K\times K}$ such that $\left[{\mathbf{P}}\,{\mathbf{u}}\right]_{m}\leq\left[{\mathbf{P}}\,{\mathbf{u}}\right]_{m+1}$, $\forall m\in[1\ldots K-1]$}
\FOR{$n=1$ \TO$N$}
\STATE{$y_{n}=\left[{\mathbf{P}}\,\mbox{{\boldmath{$\upvarphi$}}}\right]_{n}$}
%\FOR{$m=1$ \TO$N$}
\STATE{$\forall m\in[1\ldots N]:r_{m,n}=\phi(\left[{\mathbf{P}}\,\mbox{{\boldmath{$\uptau$}}}\right]_{m}-\left[{\mathbf{P}}\,\mbox{{\boldmath{$\uptau$}}}\right]_{n})$}
%\ENDFOR
\ENDFOR
\STATE{$\mbox{{\boldmath{$\upalpha$}}}=\left({\mathbf{R}}+\varepsilon\,{\mathbf{I}}\right)^{-1}\,{\mathbf{y}}$}
%\FOR{$n=1$ \TO$N$}
\STATE{$\forall n\in[1\ldots N]:e_{1,n}=\tau_{k}-\left[{\mathbf{P}}\,\mbox{{\boldmath{$\uptau$}}}\right]_{n}$}
%\ENDFOR
\STATE{$f_{{\mathrm{int}}}(\tau_{k})={\mathbf{E}}\,\mbox{{\boldmath{$\upalpha$}}}$}
\ENDFOR
\RETURN{$\{(\tau_{k},f_{{\mathrm{int}}}(\tau_{k}))\}_{k=K+1}^{M}$}
\end{algorithmic}
\caption{\label{alg: non-iterative}Non-iterative interpolation.}
\end{center}
\end{algorithm}
%%------------------------
So far, we have assumed that we have collected a batch of $M$ samples before to apply regression and interpolation, and that we estimate $\left(M-N\right)$ values jointly. This implies a long delay if $M$ is large, which is highly undesirable in practice, aggravated by the ${\mathcal{O}}(N^{3})\approx{\mathcal{O}}(M^{3})$ complexity. To reduce delay and complexity, we propose now a sample-by-sample method in which we deploy the regression and interpolation steps locally over a sliding window centered on the saturated sample $(t_{M},f_{{\mathrm{s}}}(t_{M}))$, with $M=N+1$ and $N$ small. Then, we build $S=\{(t_{n},f_{n})\}_{n=1}^{N}$ out of its $N$ nearest (anterior and posterior) unsaturated samples. While building $S$, no skipping of samples is needed for isolated saturation events, while either a recursive application of the procedure or an augmented-margin window is needed when their density is higher. The pseudo-code that corresponds to the augmented-margin approach (window length $N$, margin length $\left(M-1-N\right)$) is given in Algorithm~\ref{alg: non-iterative}.

Because we apply the sliding-window approach once per saturated sample, the average computational effort of its regression step is ${\mathcal{O}}(\rho\,N^{3})$, where $\rho={\mathrm{E}}\{\frac{M-N}{N}\}$ is the average density of saturated samples---the expected value of the saturated-to-unsaturated ratio. The cost of the sorting operation implied by Step~14 of Algorithm~\ref{alg: non-iterative} is ${\mathcal{O}}(\rho\,K\,\log K)$, with $N<K<M$. The computational effort of the interpolation step is ${\mathcal{O}}(\rho\,N)$, which is negligible. The cost of finding unsaturated samples is ${\mathcal{O}}(1)$. In the favorable case where all saturation events are isolated, the cost is finally dominated by the peak computational demand ${\mathcal{O}}(\rho\,N^{3})={\mathcal{O}}(N^{2})$.

%%%%%%%%%%%%%%%%%%%%%%%%%%%%%%%%%%%%%%%%%%%%%%%%%%%%%%%%%%%%%%%%%%%%%%%%%%%%%%%%
\subsection{Wireline {\textit{versus}} Wireless}\label{sec: Wireline versus Wireless}
%%------------------------
\begin{figure}
\begin{center}
\includegraphics[trim = 75mm 75mm 75mm 75mm,clip=true,width=3.25in]{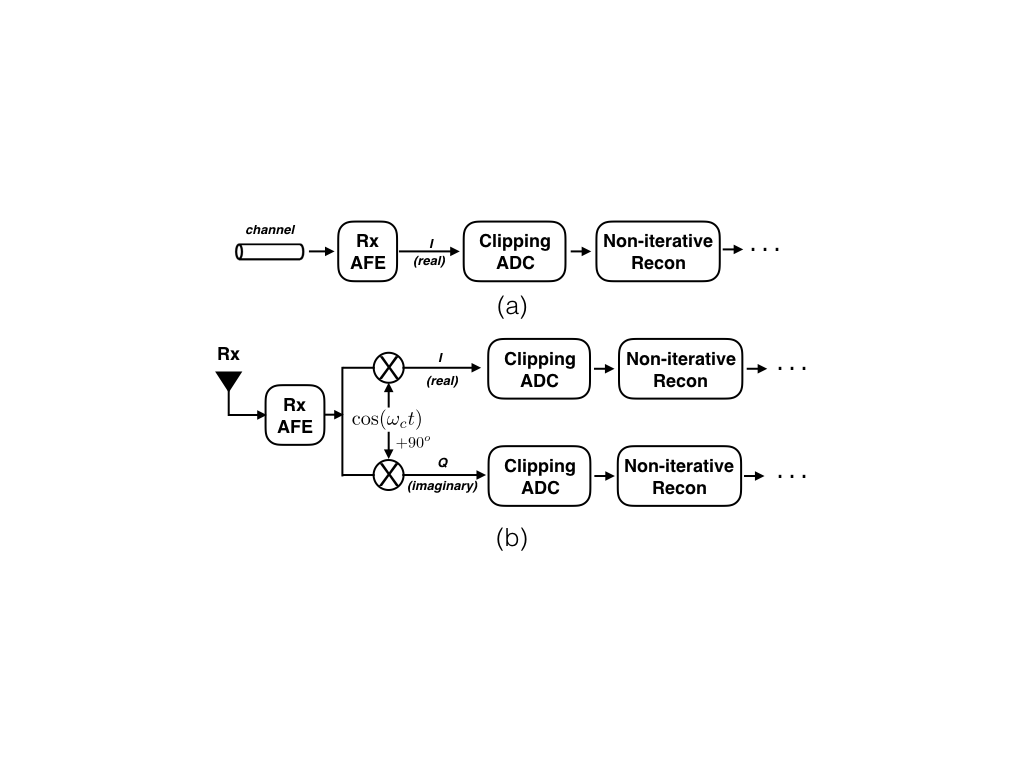}
\caption{\label{fig: OFDM configurations}Two OFDM configurations. Top: wireline. Bottom: wireless.}
\end{center}
\end{figure}
%%------------------------
In wireline applications (top part of Figure~\ref{fig: OFDM configurations}), the front-end receiver is sometimes called a discrete multi-tone transceiver. The symbols of the subcarriers, also known as Fourier coefficients, honor a Hermitian symmetry. Consequently, the temporal signal is real-valued and Algorithm~\ref{alg: non-iterative} can be applied directly.

By contrast, in wireless applications (bottom part of Figure~\ref{fig: OFDM configurations}), the symbols of the subcarriers do not follow a Hermitian symmetry. Instead, the temporal signal is split in two components transmitted separately trough the in-phase channel and the quadrature channel. Together, they represent a complex-valued time-domain signal whose real and imaginary parts are recovered by multiplication with the phase-shifted carrier frequency. But each part retains the property of being band-limited, which allows our proposed solution to be applied independently to the real and to the imaginary channel.

%%%%%%%%%%%%%%%%%%%%%%%%%%%%%%%%%%%%%%%%%%%%%%%%%%%%%%%%%%%%%%%%%%%%%%%%%%%%%%%%
\section{Results}\label{sec: Results}
We now describe our experimental setups and simulation results for five scenarios that each illustrate one aspect of our approach:
\begin{enumerate}
\item Bandpass illustrative case (Section~\ref{sec: Ideal Case}).
\item Dependence of the quality of the reconstruction on the number $N$ of nearest neighbors (Section~\ref{sec: Quality}).
\item Robustness with respect to additive Gaussian noise in wireless applications (Section~\ref{sec: Robustness}).
\item Dependence of the computational cost on the number $N$ of nearest neighbors (Section~\ref{sec: Computational Cost}).
\item Recovery of saturated samples in discrete multi-tone wireline serial-link transceivers. In this simulation, we consider realistic perturbations such as channel distortion, quantization error, and cyclic prefix (Section~\ref{sec: Wireline Communication Links}).
\end{enumerate}
Also note that the first case is bandpass, while all other cases are lowpass, with $\omega_{0}=0$ and $\omega_{1}\leq\frac{\omega}{2}$, where $\omega$ is the Nyquist rate.
The main measure of our quantitative evaluations is the bit-error-ratio (BER). We also provide the clipping ratio (CR) defined as
\begin{equation}
\label{eq: CR}{\mathrm{CR}}=\frac{\gamma}{\left\|f_{{\mathrm{s}}}\right\|_{2}},
\end{equation}
where $\left\|\cdot\right\|_{2}$ is the root-mean-square norm ({\textit{i.e.}}, the $\ell_{2}$ norm) and $\gamma$ is a threshold value.

%%%%%%%%%%%%%%%%%%%%%%%%%%%%%%%%%%%%%%%%%%%%%%%%%%%%%%%%%%%%%%%%%%%%%%%%%%%%%%%%
\subsection{Numeric Simulations}\label{sec: Numeric Simulations}

%%%%%%%%%%%%%%%%%%%%%%%%%%%%%%%%%%%%%%%%%%%%%%%%%%%%%%%%%%%%%%%%%%%%%%%%%%%%%%%%
\subsubsection{Ideal Case}\label{sec: Ideal Case}
%%------------------------
\begin{figure}
\begin{center}
% \hline
% \vline
\includegraphics[trim = 25mm 20mm 30mm 20mm,clip=true,width=3.25in]{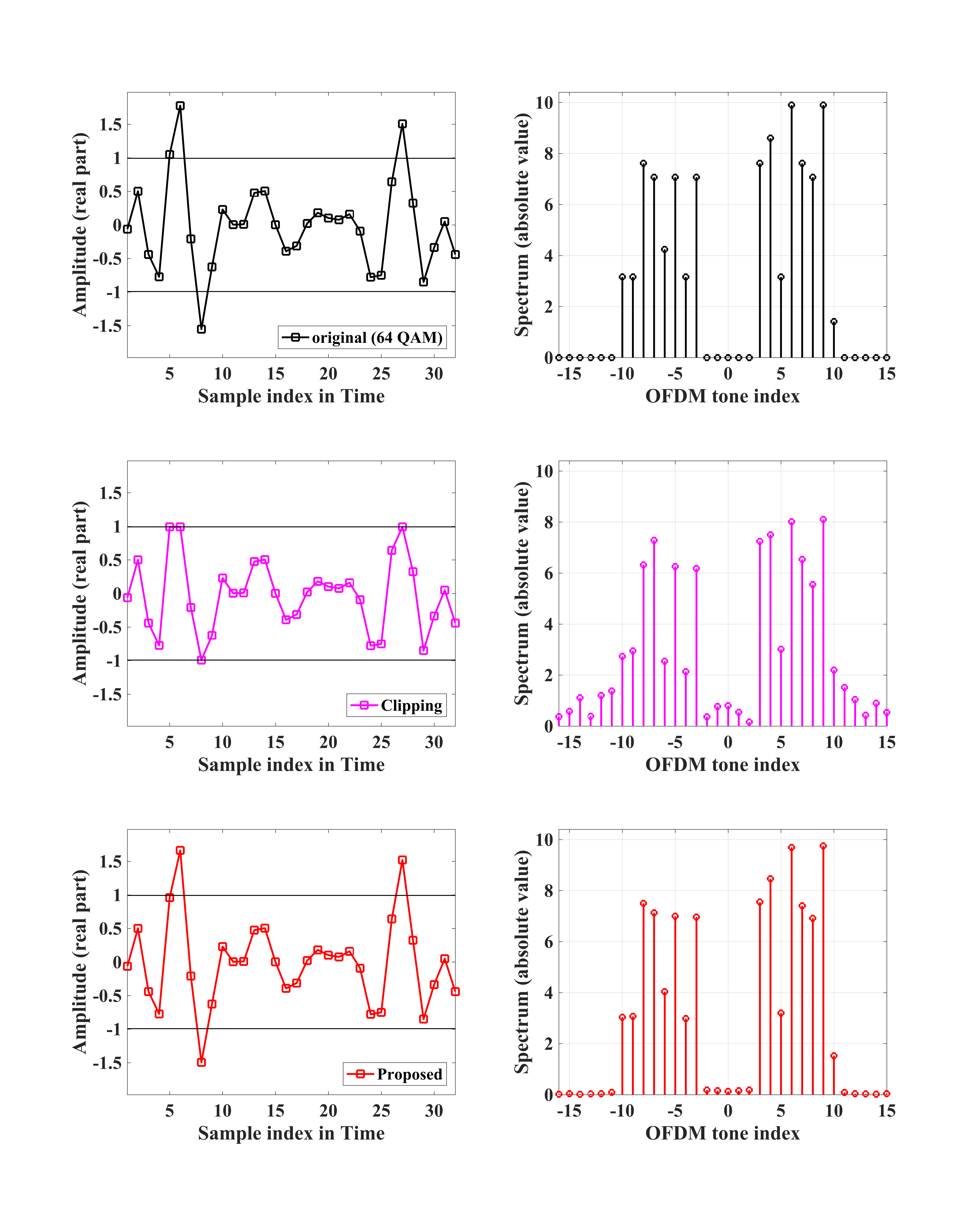}
% \vline
% \hline
\caption{\label{fig: five saturated samples}Reconstruction of five saturated OFDM values in a wireless application (complex signal). Horizontal lines: under- and over-saturation thresholds.}
\end{center}
\end{figure}
%%------------------------
We assume here that the original signal $f$ is noiseless and band-limited, which we enforce by Fourier synthesis through a $32$-tap IDFT. We simulate a wireless (complex-valued) OFDM model characterized by a zero-forced DC component, the use of sixteen subcarriers out of a total of thirty-two, and symbol encoding by $64$-QAM. We show in the top part of Figure~\ref{fig: five saturated samples} the sampled ground-truth data $f$ observed in the imaginary (Q channel) of the simulated system. Its saturated counterpart $f_{{\mathrm{s}}}$ is shown in the middle part of Figure~\ref{fig: five saturated samples}. It is apparent from $\left|\hat{f}_{{\mathrm{s}}}\right|$ that saturation produces in-band and out-of-band distortions in OFDM tones. Unless cared for, these distortions prevent the receiver from correctly retrieving symbols.

We then deploy our method with $N=8$, which compensates for saturation by minimizing the in-band and out-of-band distortions. The outcome is illustrated in the bottom part of Figure~\ref{fig: five saturated samples}. The PAPR of the original and saturated signal is $6.88$ and $4.2$, respectively. The BER of the saturated and reconstructed signals is $0.14$ and $0$ out of the transmitted bits, respectively.

%%%%%%%%%%%%%%%%%%%%%%%%%%%%%%%%%%%%%%%%%%%%%%%%%%%%%%%%%%%%%%%%%%%%%%%%%%%%%%%%
\subsubsection{Quality}\label{sec: Quality}
%%------------------------
\begin{figure}
\begin{center}
\includegraphics[trim = 0mm 0mm 0mm 0mm,clip=true,width=3.25in]{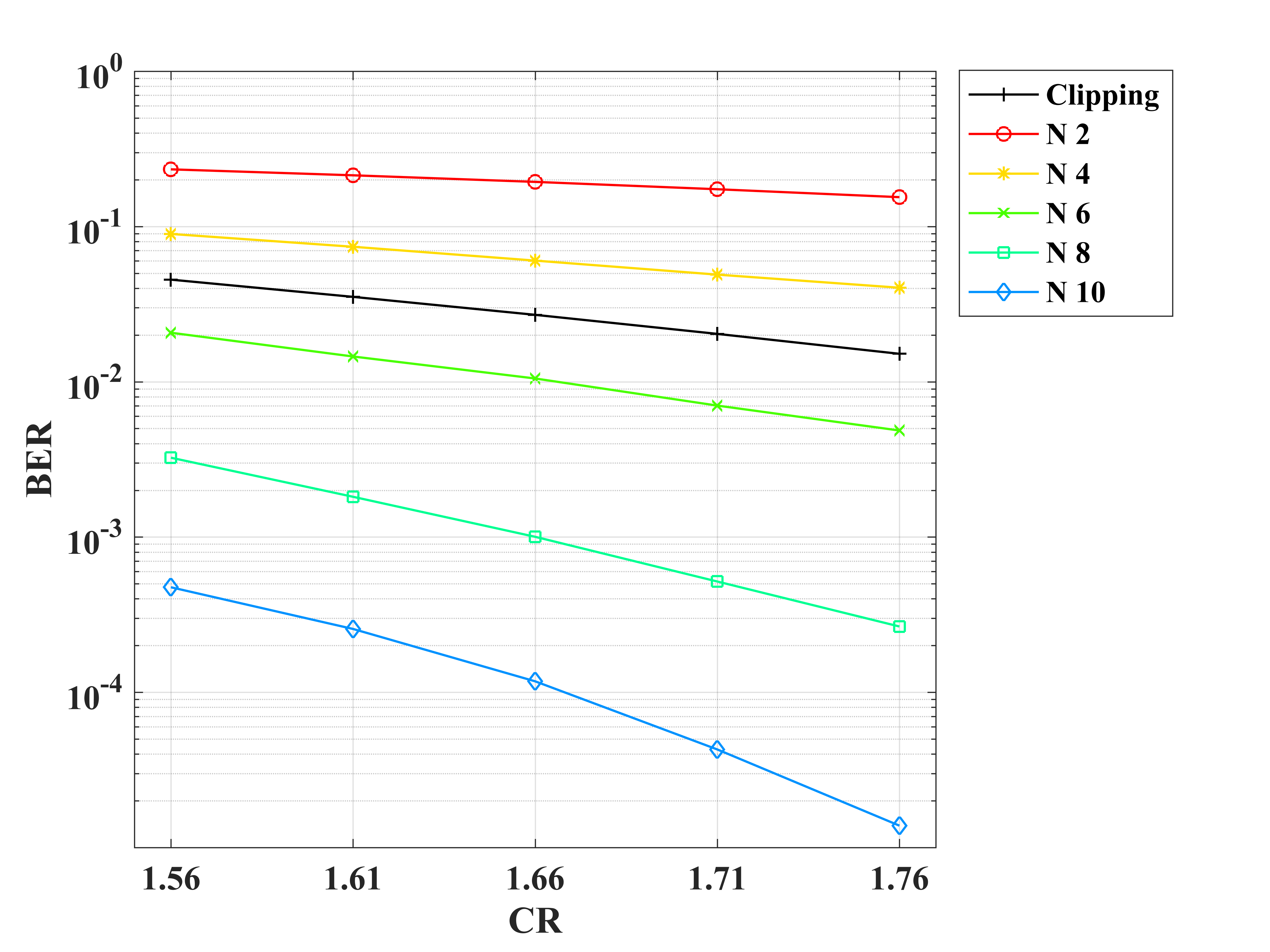}
\caption{\label{fig: wireline}Bit error rate {\textit{versus}} clipping ratio in a wireline OFDM context.}
\end{center}
\end{figure}
%%------------------------
\begin{figure}
\begin{center}
\includegraphics[trim = 0mm 0mm 0mm 0mm,clip=true,width=3.25in]{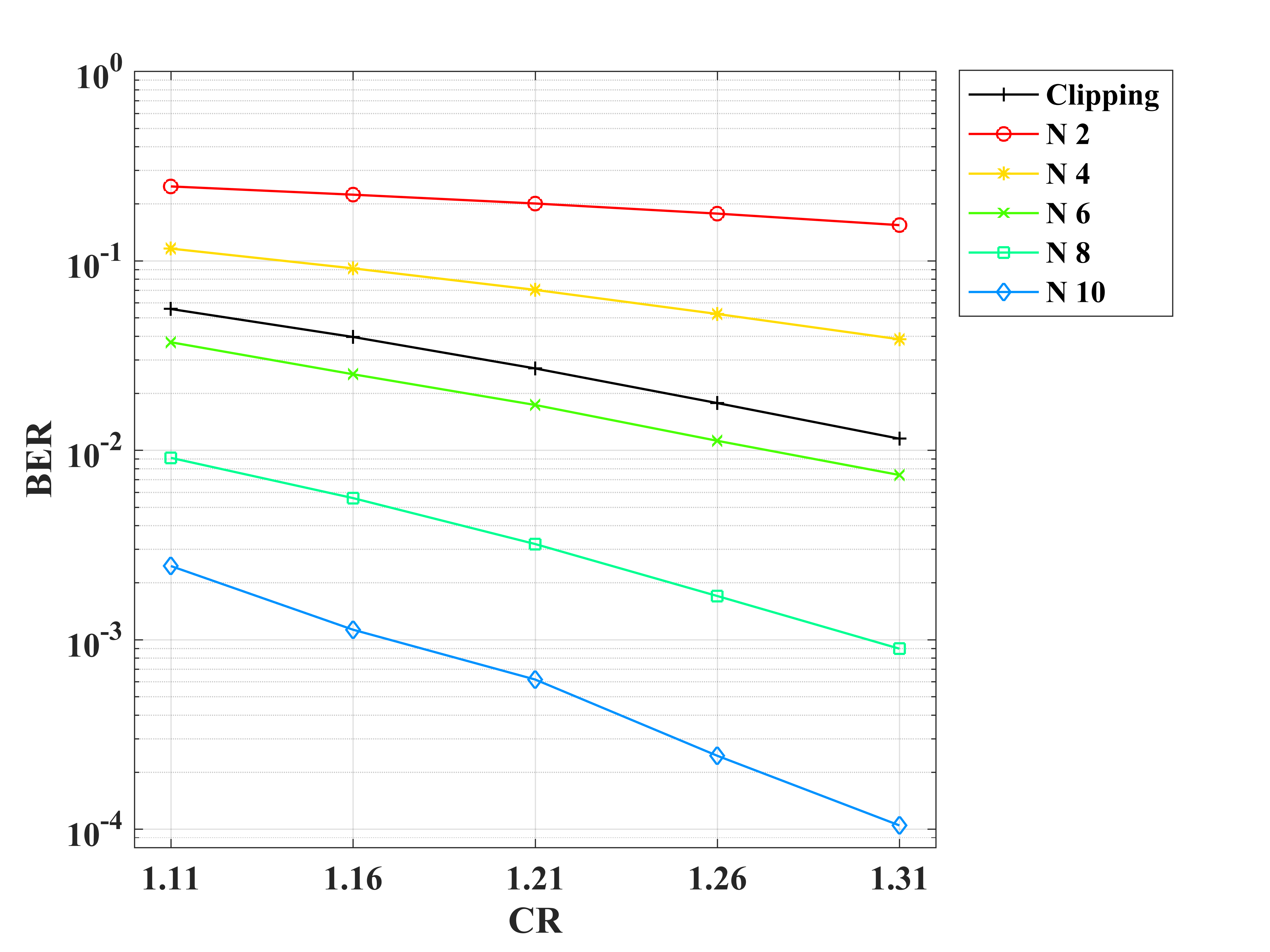}
\caption{\label{fig: wireless}Bit error rate {\textit{versus}} clipping ratio in a wireless OFDM context.}
\end{center}
\end{figure}
%%------------------------
In this section, we investigate how the BER depends on the number $N$ of unsaturated neighbors. We performed $1,\!000,\!000$ random simulations, both in a wireline and a wireless context. Each trial involved an OFDM stream based on a $32$-tap IDFT, a bandwidth of $16$ taps in the discrete domain. The DC component was maintained to zero. As we want now to focus on the sole impact of $N$, we refrained from including distortions such as insertion loss in the wireline case, additive noise, or multi-path fading in the wireless case.

The results for the wireline context are presented in Figure~\ref{fig: wireline}. There, the Hermitian symmetry reduces to $8$ the number of independent subcarriers. We observe that the proposed scheme is able to robustly estimate up to three saturated sample values (approximately, CR=1.66) with a BER less than $10^{-4}$ while using no more than $10$ neighboring unsaturated samples. It is known~\cite{frans201756} that our relatively high raw BER could be reduced much further to, say, $10^{-15}$ when combined with a forward error-correction code such as Reed-Solomon~\cite{frans201756}. Together, they satisfy the usual BER requirements of an OFDM-based high-speed point-to-point wireline serial link. When a more stringent BER target is imposed, it is enough to raise $N$. For instance, $N\geq16$ would guarantee a robust reconstruction even in the presence of five saturated samples each thirty-two samples.

The results in the wireless context are presented in Figure~\ref{fig: wireless}. We observe again that $N=10$ neighboring unsaturated samples are enough to successfully estimate up to saturations from CR 1.31. The increase in performance of the wireless over the wireline case results from the fact that the complex-valued signal of the former is split into two separate channels, thus decreasing the effective number of saturated samples per channel.

%%%%%%%%%%%%%%%%%%%%%%%%%%%%%%%%%%%%%%%%%%%%%%%%%%%%%%%%%%%%%%%%%%%%%%%%%%%%%%%%
\subsubsection{Robustness}\label{sec: Robustness}
%%------------------------
\begin{figure}
\begin{center}
\includegraphics[trim = 0mm 0mm 0mm 0mm,clip=true,width=3.25in]{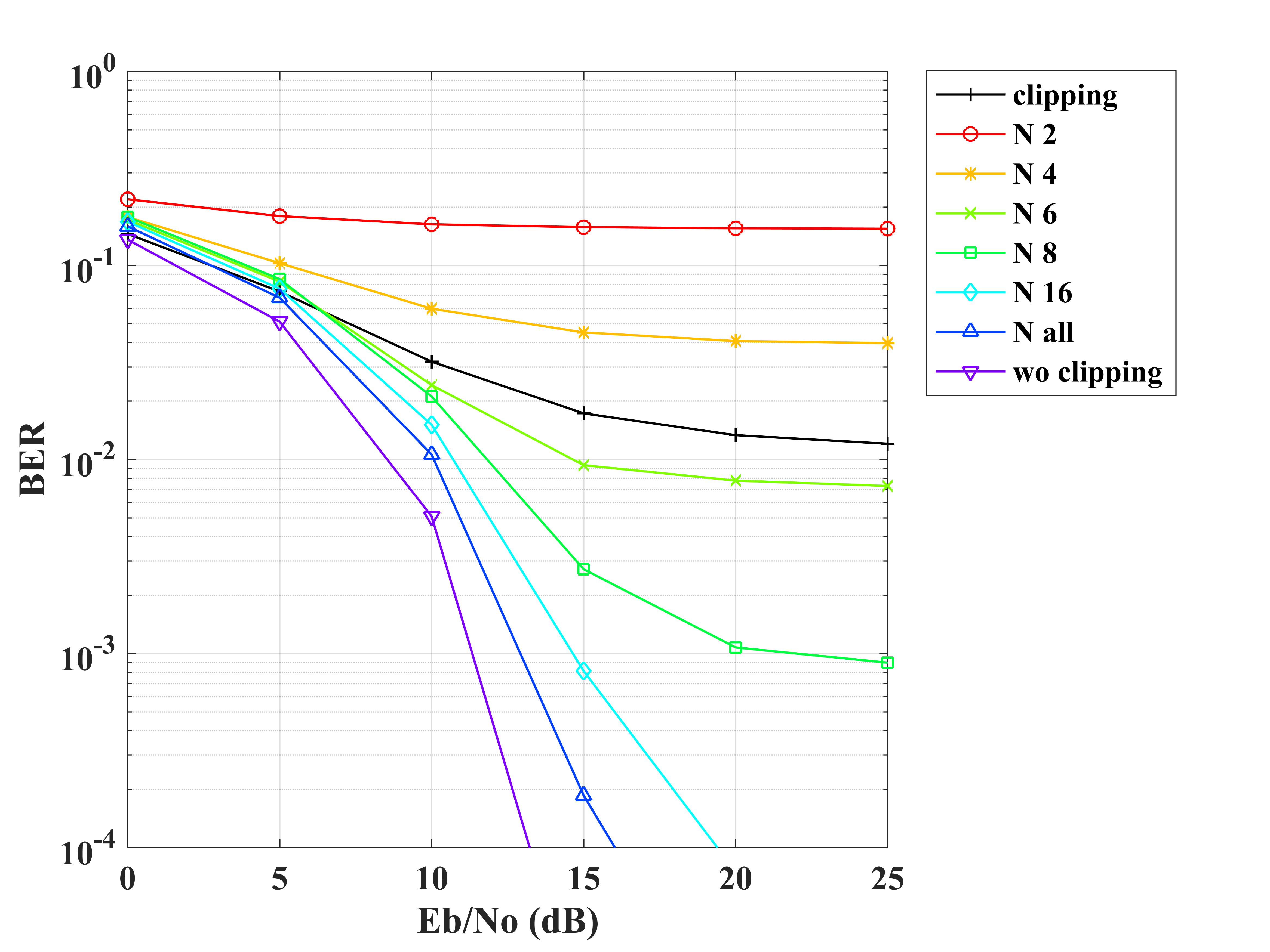}
\caption{\label{fig: robustness}Bit error rate {\textit{versus}} signal-to-noise ratio in a wireless OFDM context. The clipping ratio is fixed at ${\mathrm{CR}}=1.31$.}
\end{center}
\end{figure}
%%------------------------
It is typical for wireless long-distance communications to be corrupted by several sources of noise. We simulate now their impact by the channel-wise addition of white Gaussian noise, with a controlled signal-to-noise ratio. Like in our previous experiments, a $32$-tap signal is used for the forward/backward model of OFDM, the bandwidth is half of the total length of the signal, and $64$-QAM is used for symbol encoding; the DC component is zeroed. We report the resulting BER in Figure~\ref{fig: robustness}. Once again, we observe that $N=8$ leads to a performance that is very similar to what would be obtained in the absence of saturation and confirms the robustness of our method in the presence of noise.

%%%%%%%%%%%%%%%%%%%%%%%%%%%%%%%%%%%%%%%%%%%%%%%%%%%%%%%%%%%%%%%%%%%%%%%%%%%%%%%%
\subsubsection{Computational Cost}\label{sec: Computational Cost}
%%------------------------
\begin{figure}
\begin{center}
\includegraphics[width=3.25in]{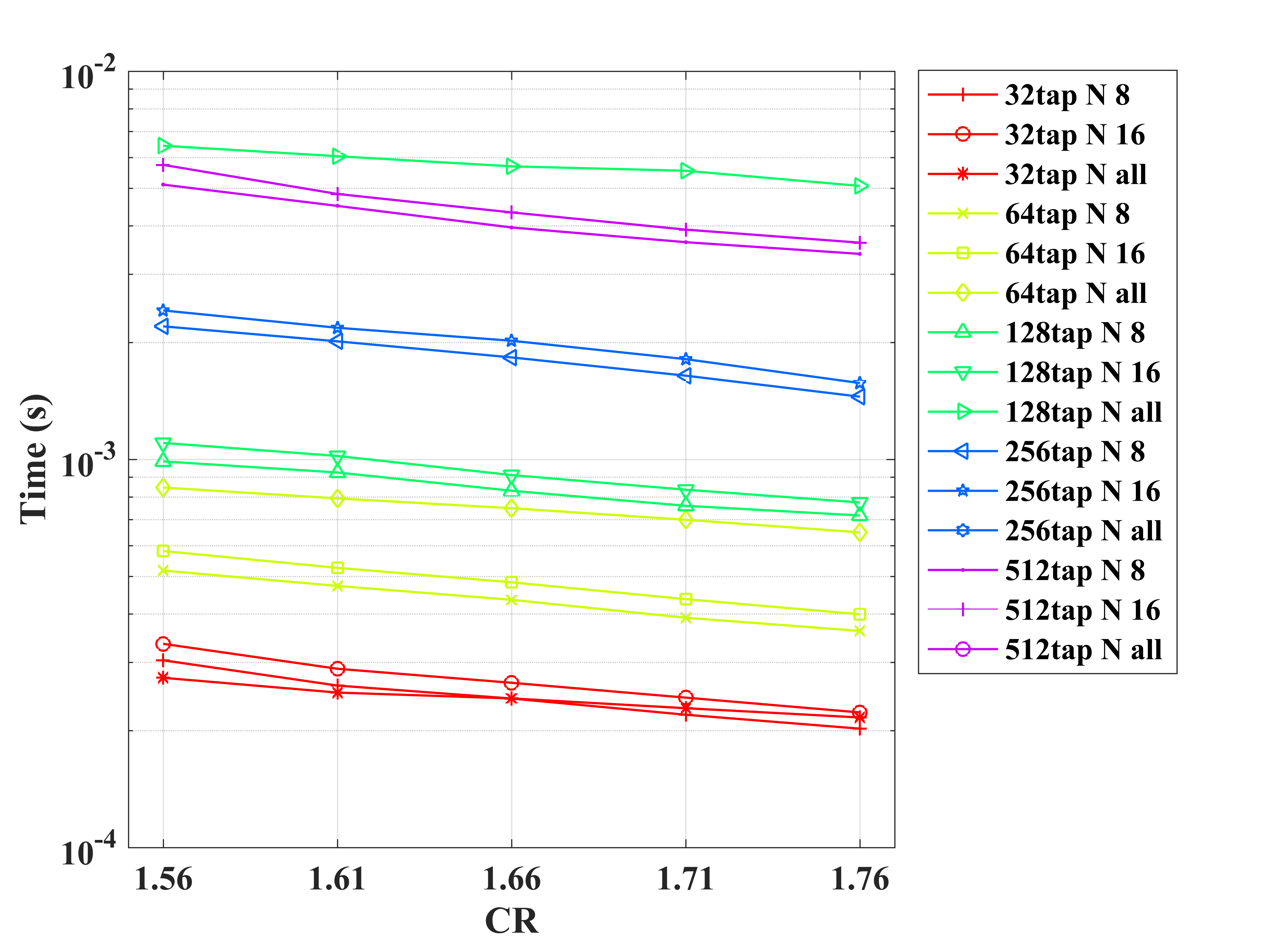}
\caption{\label{fig: computational cost}Computation time {\textit{versus}} clipping ratio for a single sliding window in a wireline OFDM context.}
\end{center}
\end{figure}
In Figure~\ref{fig: computational cost}, we report the computation times observed for $32$- to $512$-tap signals and various clipping ratios. We see that they correlate weakly with CR while, in first approximation, they increase linearly with the number of taps. We attribute this dependence to the combined cost of sorting and matrix inversion, given that the number of values to estimate (the number of saturated samples) increases in proportion of the number $M$ of taps.

Time savings can be easily obtained if the sample locations sit on a grid. Then, the entries of the matrix ${\mathbf{R}}$ to invert in~(\ref{eq: regression}) and the matrix ${\mathbf{E}}$ to apply in~(\ref{eq: interpolation}) can all be precomputed. When the saturation events are either isolated or in simple configurations, it is convenient to store pre-inverted matrices in a configuration-dependent lookup table. The computational cost is then dominated by the sorting step needed to find the neighbors nearest to each saturation event.

%\fbox{\parbox{3in}{{\textbf{PT: I failed to understand what you mean.}} Therefore, the total complexity using the truncated basis is reduced from by O(Pr3) from O(M3) (r ? M, (i.e. lines from 128 FFT at 1.66 CR in Fig. 9, CR is given in Eq. (16)), r = 8 and r = 128?P cases takes 8?10?4 and 6?10?3 seconds, respectively. Thanks to this approximation, we can significantly reduce the computation time in the practice as shown in Fig. 9.}}

%%%%%%%%%%%%%%%%%%%%%%%%%%%%%%%%%%%%%%%%%%%%%%%%%%%%%%%%%%%%%%%%%%%%%%%%%%%%%%%%
\subsection{Wireline Communication Links}\label{sec: Wireline Communication Links}
%%------------------------
\begin{figure}
\begin{center}
\includegraphics[width=3.25in]{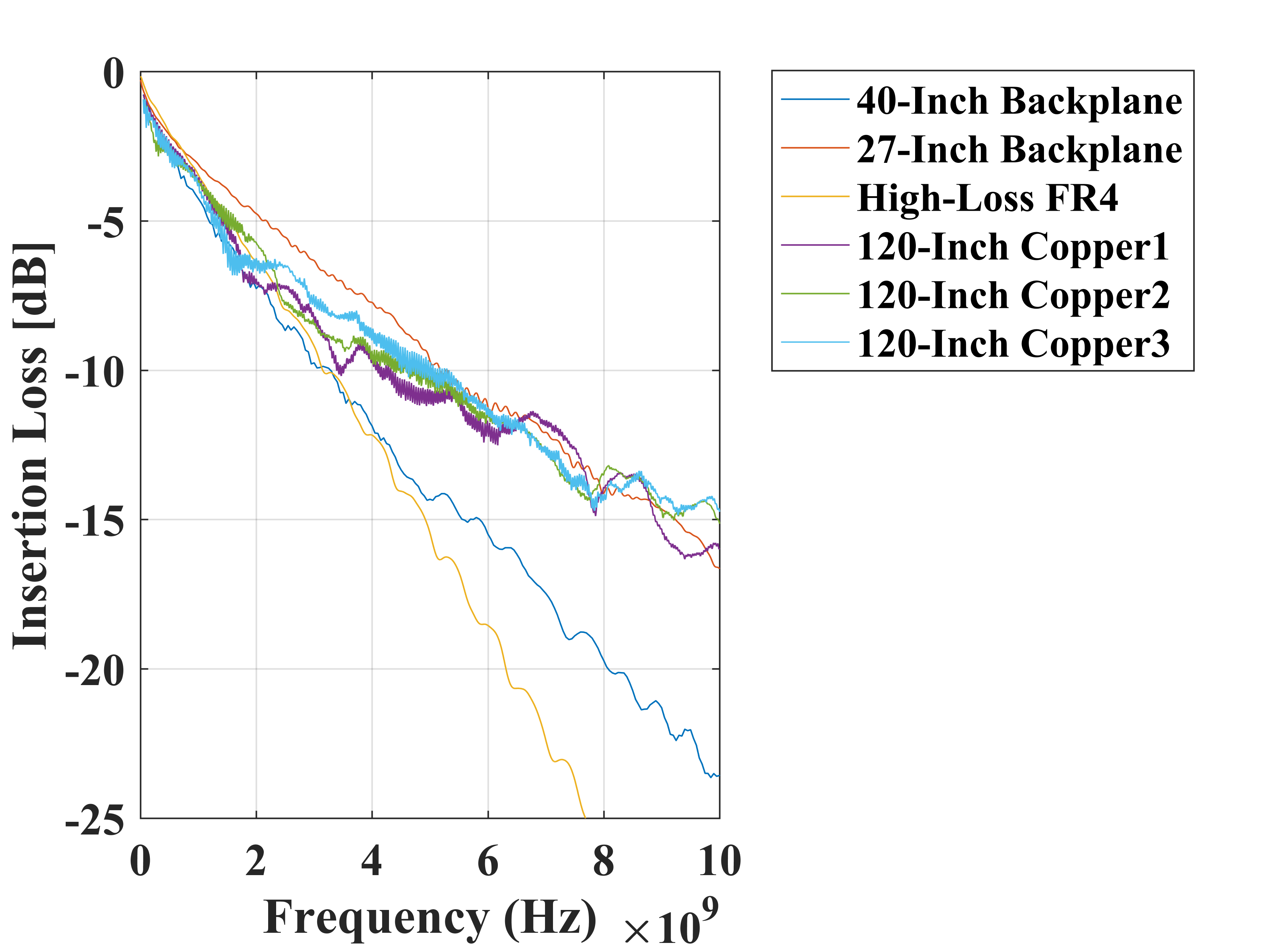}
\caption{\label{fig: insertion loss}Frequency-dependent insertion loss.}
\end{center}
\end{figure}
%%------------------------
\begin{figure}
\begin{center}
\includegraphics[width=3.25in]{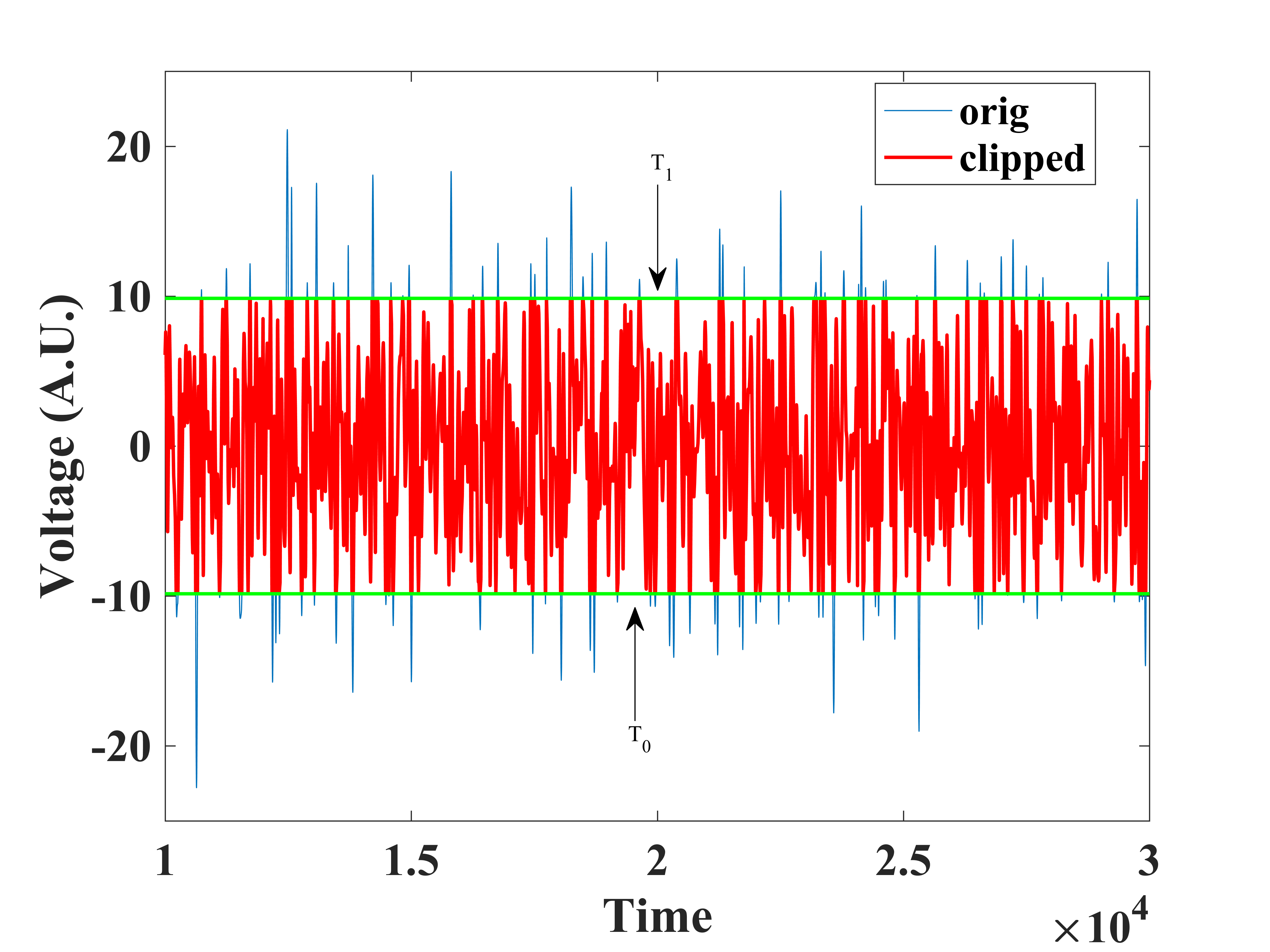}
\caption{\label{fig: ADC input}OFDM signals on the receiver frontend. Thin curve: original unsaturated. Thick curve: saturated (clipped). The clipping ratio is ${\mathrm{CR}}=2.03$.}
\end{center}
\end{figure}
%%------------------------
\begin{figure}
\begin{center}
\includegraphics[width=3.25in]{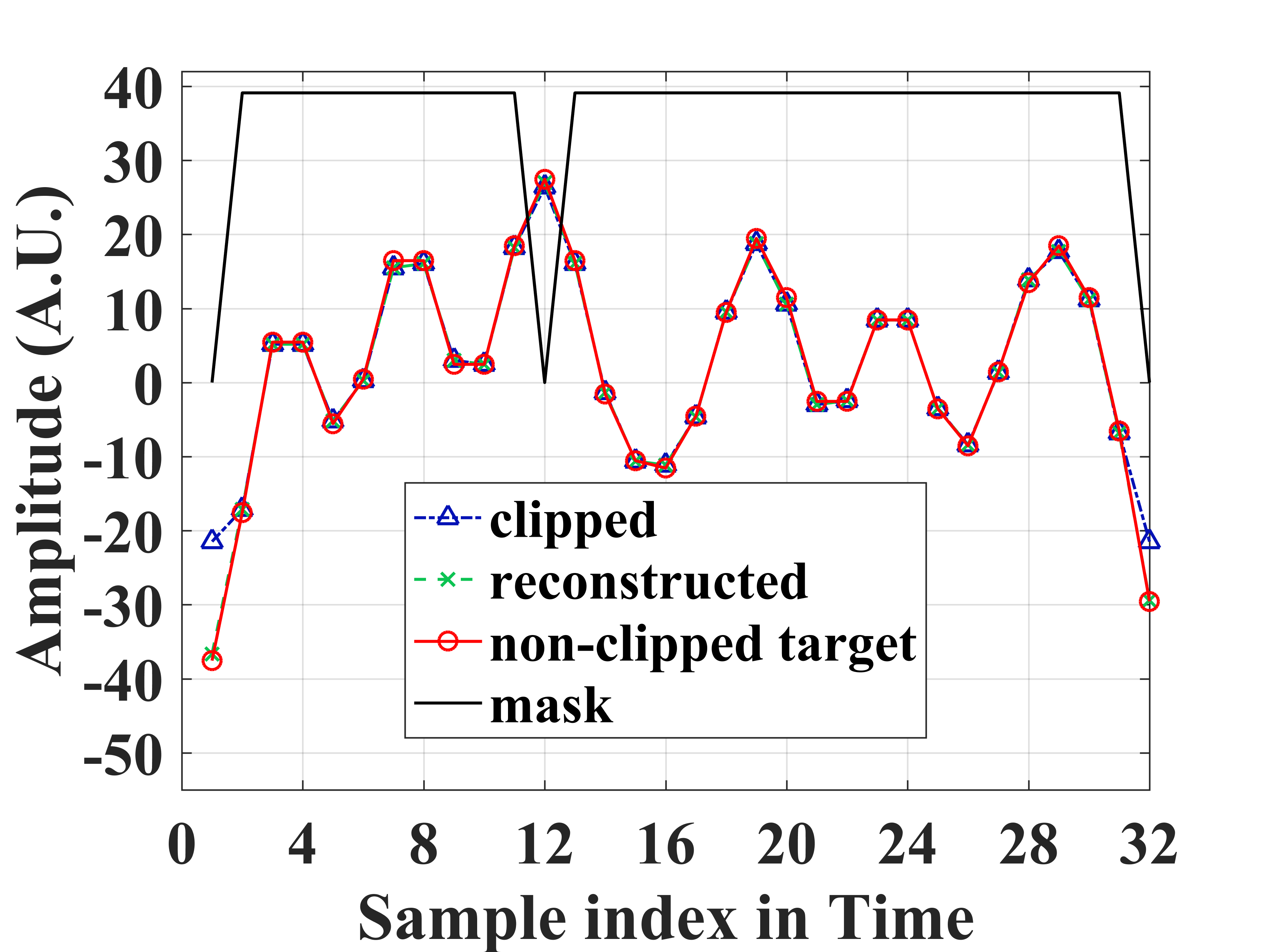}
\caption{\label{fig: reconstruction}Reconstruction of an OFDM symbol.}
\end{center}
\end{figure}
%%------------------------
\begin{figure}
\begin{center}
\includegraphics[width=3.25in]{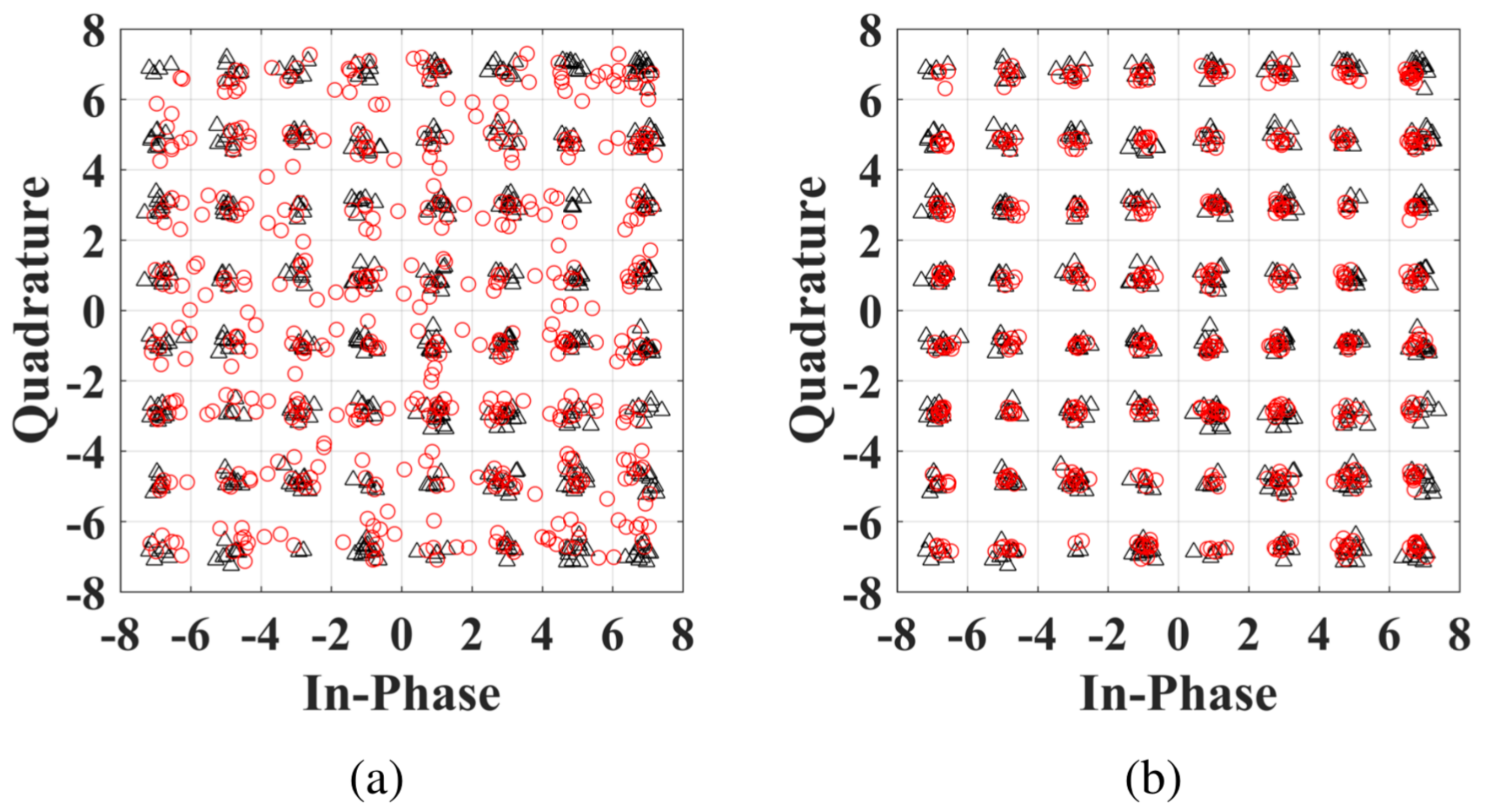}
\caption{\label{fig: constellation diagrams}Constellation diagrams. Triangles: ground-truth before saturation. Circles: data degraded by saturation. Left: data processed without compensation for saturation. Right: proposed method.}
\end{center}
\end{figure}
%%------------------------
In electrical wireline transceiver systems, the insertion loss of the transmission line (channel) is the strongest source of noise when the channel is lossy. We illustrate in Figure~\ref{fig: insertion loss} the actual frequency response of real four-lane $100\,{\mathrm{Gb}}\,{\mathrm{s}}^{-1}$ $40$-inch/$27$-inch backplane channels, a long-reach lossy FR-4 channel, and three different $120$-inch twin-axial copper cables.

When the OFDM signal is transmitted through such channels, the worst-case PAPR grows as the signal gets farther away from the emitter. In particular, while the non-zero spectral components are spread evenly over the entire frequency range, the DC component might still be responsible for large amplitudes because it undergoes much less attenuation that the high-frequency components. At the input of the receiver, this results in a wide dynamic range that is difficult to quantize since it is desired to faithfully represent, on one hand, the small signal variations that encode the information carried by high-frequency components and, on the other hand, the large variations related to the low-frequency components. For a finite-range linear quantizer, the generally accepted tradeoff is to tolerate some saturation in return for a finer representation of the least-significant bits. This mechanism explains why the reduction of the PAPR of an OFDM signal is of high interest.

Figure~\ref{fig: ADC input} contains a snapshot of the OFDM signal at the receiver frontend, before and after the ADC. Out of $32$, this signal is made of $7$ low-frequency OFDM carriers (including DC); the cyclic prefix length is chosen to be $4$. The constellation is fixed to $64$-QAM for all sub-channels. The effective number of bits of the ADC is assumed to be $6.5$, which means that the signal-to-noise and distortion ratio is about $41\,{\mathrm{dB}}$, for a bandwidth of up to $6\,{\mathrm{GHz}}$. The ADC sampling rate is $24\,{\mathrm{GHz}}$ for a final $28\,{\mathrm{Gb}}\,{\mathrm{s}}^{-1}$ data rate. The $27$-inch backplane channel exhibits an attenuation of about $11\,{\mathrm{dB}}$ due to the transmission medium.

We provide in Figure~\ref{fig: reconstruction} a simulation of the various signals that play a role in our proposed method. Virtual samples of the unspoiled, continuously defined signal (solid line) that enters the ADC are shown with circles; these samples are not accessible in practice. What is accessible is their saturated ({\textit{i.e.}}, clipped) and quantized version (mixed line and triangles). The ADC hardware also provides a mask signal that encodes the saturation events (top solid line without markers). The saturation-compensated reconstructed signal ($N=8$) we propose is shown with a dashed line and cross markers. It follows closely the input signal.

Constellation diagrams after frequency-domain equalization for channel inversion are shown in Figure~\ref{fig: constellation diagrams}. It can be observed visually that the method we propose to recover estimated values of the true samples leads to a significant improvement of the SNR of the recovered signal.

%%%%%%%%%%%%%%%%%%%%%%%%%%%%%%%%%%%%%%%%%%%%%%%%%%%%%%%%%%%%%%%%%%%%%%%%%%%%%%%%
\section{Conclusion}\label{sec: Conclusion}
We have proposed a direct, non-iterative method to recover value estimates for those samples of OFDM signals that are saturated. The non-iterative nature of our algorithm dramatically reduces its implementation cost, while we showed through simulations that our solution performs well and offers a strong mechanism to compensate for information losses related to saturation.

Our main prior assumption is that the original signal is band-limited. We have developed a theoretical framework that justifies our construction under this specific assumption. This theory extends the scope of our proposed solution beyond the sole application of reconstructing OFDM saturated samples.

Although the most expensive computations of our method are related to a matrix-inversion step, we were able to show that the handling of a small $\left(8\times8\right)$ matrix is often sufficient in realistic wireline/wireless digital-communication systems. As the matrix components do not depend on data, future versions of the the algorithm will also rely on lookup tables so that the matrix-inversion step can be bypassed altogether.

The low computing complexity of our proposal shows potential for an efficient hardware implementation, possibly running at hundreds of MHz or even in the GHz domain once implemented on silicon as an application-specific integrated circuit (ASIC), aiming at real-time operations.

\bibliographystyle{IEEEtran}
% \clearpage
%\bibliography{biblio_kh}


\end{document}